\documentclass[iop]{emulateapj}
\usepackage{epsfig}
\usepackage{amsmath}
\usepackage{natbib}
\usepackage{color}
\usepackage{psfrag}
\usepackage{graphicx}
\usepackage{subfigure}

\newcommand{\Msun}{\>{\rm M_{\odot}}}   
\newcommand{\kpch}{\>{h^{-1}{\rm kpc}}}
\newcommand{\mpch}{\>h^{-1}{\rm {Mpc}}}

\newcommand{\Msunh}{\>h^{-1}\rm M_\odot}    
\def \etal {et~al.~}    
\def\apj{ApJ} 
\def\mnras{MNRAS} 
\def\apjl{ApJL} 
\def\aap{A\&A} 
\def \prd  {PhRvD}

\begin{document}

\title{The effect of Warm Dark Matter on galaxy properties: constraints from the stellar mass function and the Tully-Fisher relation}

\author{Xi Kang$^{1}$, Andrea. V. Macci\`o$^{2}$ and Aaron A. Dutton$^2$}
\affil{$^1$Purple Mountain Observatory, the Partner Group of MPI f\"{u}r Astronomie, 2 West Beijing 
        Road, Nanjing 210008, China\\
        $^2$Max-Planck-Institut f\"ur Astronomie, K\"onigstuhl 17, 69117, Heidelberg, Germany}


\begin{abstract}
  In this paper we combine high resolution N-body simulations with a
  semi analytical model of galaxy formation to study the effects of a
  possible Warm Dark Matter (WDM) component on the observable
  properties of galaxies.  We compare three WDM models with a dark
  matter mass of $0.5$, $0.75$ and $2.0$ keV, with the standard Cold
  Dark Matter case.  For a fixed set of parameters describing the
  baryonic physics the WDM models predict less galaxies at low
  (stellar) masses, as expected due to the suppression of power on
  small scales, while no substantial difference is found at the high
  mass end.
  However these differences in the stellar mass function, vanish when
  different set of parameters are used to describe the (largely
  unknown) galaxy formation processes.
  We show that is possible to break this degeneracy between DM
  properties and the parameterization of baryonic physics by combining
  observations on the stellar mass function with the Tully-Fisher
  relation (the relation between stellar mass and the rotation
  velocity at large galactic radii as probed by resolved HI rotation
  curves).
  WDM models with a {\it too warm} candidate ($m_{\nu}<0.75$ keV)
  cannot simultaneously reproduce the stellar mass function and the
  Tully-Fisher relation.
  We conclude that accurate measurements of the galaxy stellar mass
  function and the link between galaxies and dark matter haloes down
  to the very low-mass end can give very tight constraints on the
  nature of DM candidates.
\end{abstract}




\section{Introduction}
\label{sec:intro}

The Cold Dark Matter (CDM) paradigm successfully describes the
formation of large-scale structure in the Universe (e.g., Springel et
al. 2006).  On small scales, the CDM model faces inconsistencies with
observations.  For example, CDM model over-predicts the number of
satellites in the Milky Way (e.g. Klypin et al. 1999), predicts cuspy
density profile for the dark matter halo (e.g.  Moore et al.  1999),
and a too large kinematics for massive satellites (Boylan-Kolchin et
al.  2011).

These small-scale problems in CDM all arise from the comparison of
pure N-body simulations with observational data.  Recently it has been
shown that some of those problems could be alleviate by the inclusion
of baryonic physics, due to its back-reaction on the properties of
dark matter haloes.
For example, UV background, reionization and supernova feedback can
act together to suppress the formation of dwarf satellites in the
Milky Way (e.g. Bullock et al 2001, Macci\`o et al. 2010), the inner
halo density profile becomes more flat when star formation and
feedback are included (e.g. Mashchenko et al. 2008, Pontzen \&
Governato 2012, Macci\`o et al. 2012). However, these baryonic effects
are difficult to model and still strongly rely on simple
parameterization for complex physical processes (the so called sub-grid
physics)

Another possible solution is to suppress the excess of power at small
scales in the Dark Matter distribution by changing the properties of
the dark matter candidate.  Among many of these different approaches,
the Warm Dark Matter (WDM) model is the most intriguing as it
naturally preserves the large-scale structure of the CDM model (Alam
et al.2002; Zentner \& Bullock 2002).
Many authors have shown that the WDM model can relax some tensions
between observations and theoretical predictions using pure N-body
simulations (Colin et al. 2000, Bode \etal 2001, Gotz \& Sommer-Larsen 2002,
Knebe \etal 2008, Colin \etal 2008, Tikhonov et al 2009, Lovell et al.2012, 
Macci\`o \etal 2012b).

One possible WDM candidate is a sterile neutrino which exhibit a
significant primordial velocity distribution and thus damp primordial
inhomogeneities on small scales (e.g. Hansen \etal 2002, Abazajian \&
Koushiappas 2006, Boyarsky et al. 2009).
Limits on the mass of dark matter particles can be obtained from
several astrophysical observations: one of the most powerful tool for
constraining the matter power spectrum are Lyman-$\alpha$ forest
observations (neutral hydrogen absorption in the spectra of distant
quasars, Narayanan \etal 2000, Viel \etal 2005, Seljack \etal 2006).
Lyman-$\alpha$ observations allows the possibility to studying the
power spectrum down to small scales and over a large range of
redshifts.  Current observations set a lower limit of $m_{\nu} \approx
1$ keV (Viel \etal 2008).

These limits have been confirmed by other observations based on
different methods like QSO lensing (Miranda \& Macci\`o 2007),
luminosity function of high redshift QSOs (Song \& Lee 2009), dwarf
galaxies in the Local Volume (Zavala \etal 2009), the size of
(min)-voids around the Local Group (Tikhovov \etal 2009) and Milky Way
satellites (Macci\`o \& Fontanot 2010).

Up to now, most studies based on numerical simulations have focused
their attention on the internal structure and kinematics of nearby
galaxies in the Warm Dark Matter model, while very few have explored
the effect that a warm candidate will have on the properties of the
general galaxy population, such as the luminosity function, the
stellar to halo mass function and so on.

A notable exception is the recent paper by Menci et al. (2012),
although not directly based on N-body simulations in WDM.  In their
study the authors combined a Semi Analytical Model for galaxy
formation with the Extended Press and Shechter formalism (e.g.
Cole \etal 2000, Lacey \& Cole 1993) to obtain the properties of galaxy in a WDM
model. They found that the WDM model can produce a more flat slope for
the faint end of the luminosity function in better agreement with the
observations.  They also argue that other galaxy properties better
match observations when a warm candidate is used instead than a cold
one.
While these effects are certainly true, the adoption of a Warm Dark
Matter like spectrum has also strong influence on the ratio between
dark matter and stellar mass in galaxies. This ratio that can be
constrained by several dynamical mass estimators such as the
Tully-Fisher relation (Tully \& Fisher 1977), satellites kinematics,
and weak gravitational lensing.

In this paper we combine high resolution N-body simulations for
 the standard CDM scenario and three WDM candidates
$m_{\nu}=2,0.75,0.5$ keV, with the Semi Analytical Models first
presented in Kang et al. (2005), and then successively expanded and
improved in Kang \& van den Bosch (2008) and Kang et al. (2012).  Our
aim is to study the impact of different WDM models on galaxy
properties and use different observations to try to disentangle the
effects of a warm candidate from the different parameterization of the
baryonic physics adopted in the SAM.  The final goal is to
constrain the properties of the dark matter candidate.

The paper is organized as follows: We introduce the simulations and
model in \S2, show the model predictions in comparison to observations
in \S3, and conclude in \S4.

\begin{figure*}
 \centerline{\psfig{figure=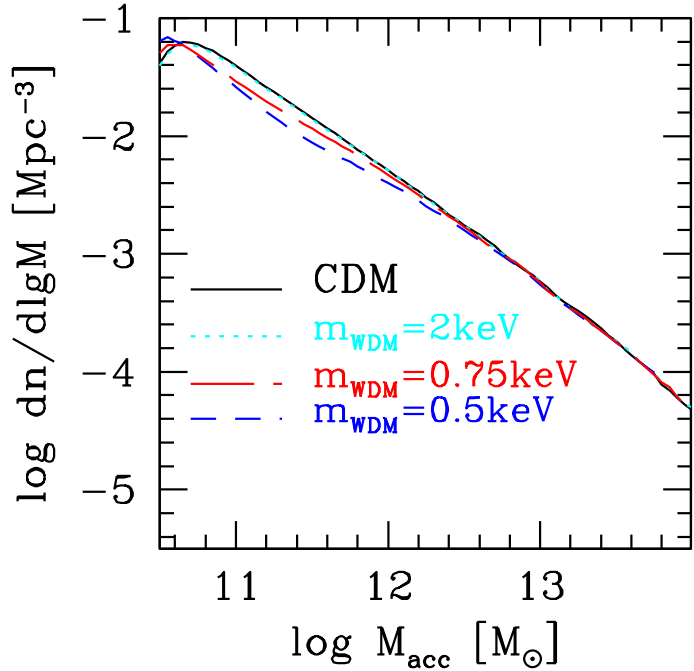,width=0.5\textwidth}}
 \caption{The halo
   mass  functions from the CDM and WDM simulations. Here the mass functions are slightly different from the usual ones, as we have included the satellite haloes, for which we use the mass at the time of accretions. }
\label{fig:HMFs}
\end{figure*}

\begin{figure*}
\centerline{\psfig{figure=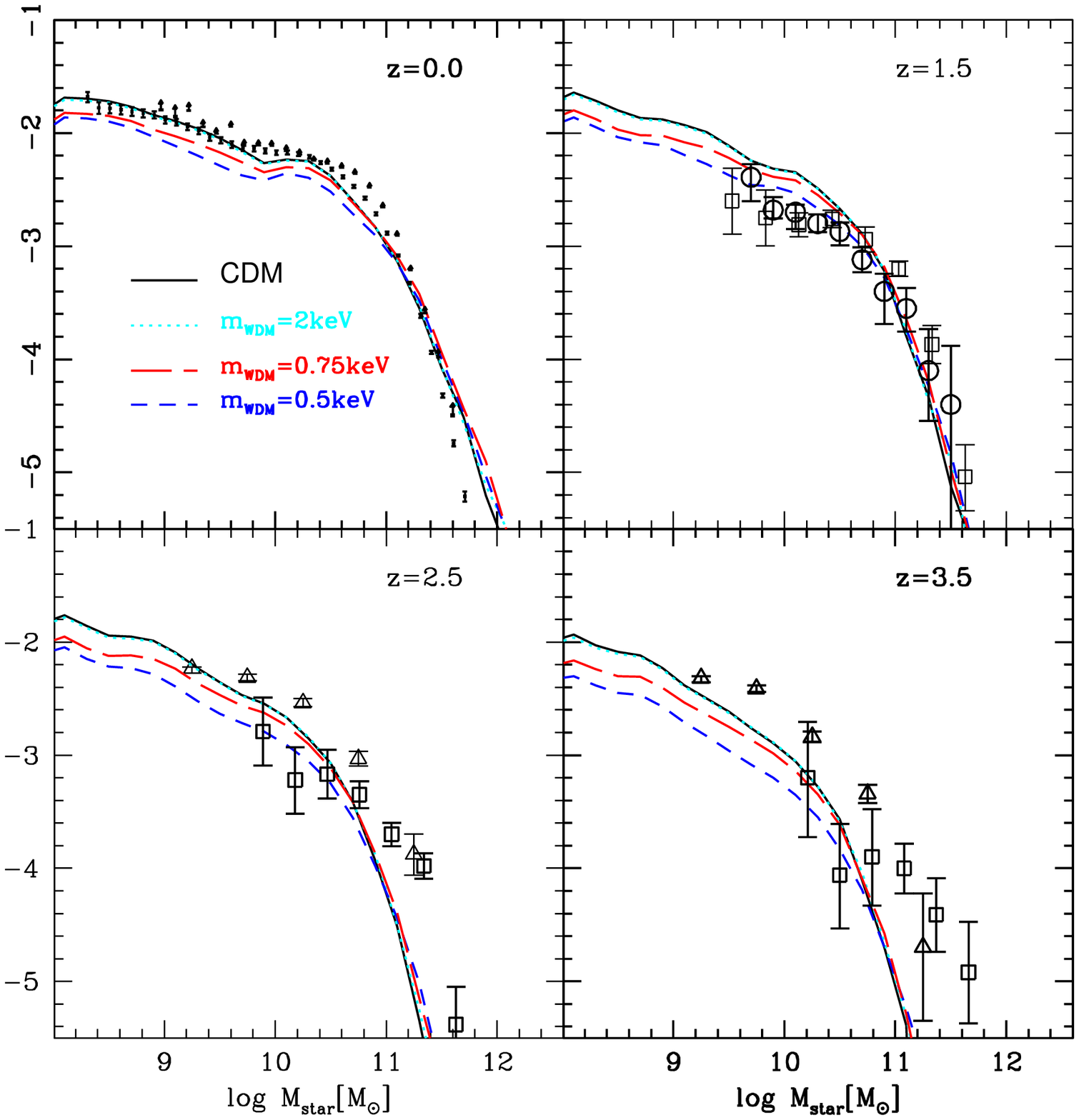,width=0.95\textwidth}}
\caption{The galaxy stellar mass functions.
lines are from our model  with $f_{\rm c}=0.15$ (see
   text). The $z=0$ data points are from Li \& White (2009) and Bell
   et al.  (2003), and high-z data points are from Marchesini et
   al. (2009), Fontana et al. (2006).}
 \label{fig:SMFs}
\end{figure*}

\section{N-body simulations and galaxy formation model}
\label{sec:method}

In the WDM model, the velocity dispersion of dark matter particles
defines a characteristic free-streaming scale, below which the
perturbation will be suppressed.  The free-streaming scale can be
described as (Bode \etal 2001),

\begin{equation}
  r_{\rm fs}=0.11 \left( \frac{\Omega_{\rm m}h^{2}}{0.15}\right )^{1/3} \left (\frac{m_{\rm \nu}}{ {\rm keV}}\right)^{-4/3} {\rm Mpc}
\end{equation}

The lighter the particle mass is, the larger the scale below which
perturbations will be suppressed. It also possible to define a
free-streaming mass scale as, $M_{\rm fs}=4/3\pi\rho_c(r_{\rm fs}/2)^3$
(where $\rho_c$ is the critical density of the Universe), and the initial density
perturbations are practically erased on masses below $M_{\rm fs}$.
The impact of the free-streaming on the power spectrum can be
described using a modified transfer function as suggested in Bode et
al. (2001):
\begin{equation}
T(k)=\left(\frac{P_{\rm Lin}^{\rm WDM}}{P_{\rm Lin}^{\rm CDM}}\right)^{1/2} = (1+(\alpha k)^{2\mu})^{-5/\mu}.
\end{equation}
Where $P_{\rm Lin}^{\rm WDM}$ and $P_{\rm Lin}^{\rm CDM}$ are the
linear power spectrum in the WDM and CDM models.  Viel et al.(2005)
using a Boltzmann simulation code found that $\mu=1.2$, and $\alpha$
can be directly related to the mass of the warm dark matter particles
and the matter density parameter in the following way:
\begin{equation}
\alpha = 0.049  \left ( {m_{\rm \nu} \over  \rm{1 keV}}  \right )^{-1.11}   \left ( { \Omega_{\rm \nu}  
\over 0.25 }\right )^{0.11}   \left ( { h \over 0.7} \right )^{1.22} \mpch.   
\end{equation}

We used the above equations, to derive the initial power spectrum for
our WDM models, and take it as input for generating the initial
conditions of our N-body simulation.

The cosmological parameters are set in agreement with those derived
from the WMAP7 data release (Komatsu \etal 2011), namely:
$\Omega_{\Lambda}$=0.73, $\Omega_{\rm m}$=0.27, $\Omega_{\rm
  b}$=0.044, $h=0.7$ and $\sigma_8=0.81$.  The simulations have been
run using the {\sc gadget-2} code (Springel 2005) in a box of
$L=200$ Mpc/h using $1024^{3}$ particles. This sets our mass
resolution to $5.5 \times 10^8 \Msunh$ and the softening to $4.5
\kpch$.  We run three different WDM models for a thermal candidate
with particle mass of $0.5$, $0.75$ and $2.0$ keV respectively.
Finally, readers are refereed to Kang \etal (2005) for details on the
halo finder and how we construct merger trees from the simulations.

The Semi Analytical Model (SAM) of galaxy formation we employ in this
study is based on the model of Kang \etal (2005, 2006). This SAM
self-consistently models the physical processes governing stellar mass
evolution, such as gas cooling, star formation, supernova and active
galactic nucleus (AGN) feedback.  The latest incarnation of this model
had been recently presented in Kang \etal (2012) and we refer the
reader to this paper for more details.  As in Kang et al. (2012), we
slightly modified the cooling rate of low-mass haloes as,
\begin{equation}
\dot{M}_{\rm cool} = f_{\rm c} \times  m_{\rm hot}/t_{\rm dyn}.
\label{eq:cooling}
\end{equation}
where $f_{\rm c}$ is the cooling factor.  As described in Kang \etal
(2012), this formula can be interpreted as an effective rate for gas
cooling, that takes into account the gas outflow due to reheating by
supernova feedback.

Kang \etal (2012) has shown that a low value for $f_{\rm c}$ is needed
to better reproduce the slope of the low-mass end of the galaxy
stellar mass function in the CDM model.  We will show later in \S3
that for the WDM models, due to lower power on small scales, this
cooling factor $f_{\rm c}$ has to be increased to match the local
stellar mass function.

\section{Results}

In the WDM model, the formation of haloes with mass below the
free-streaming scales is suppressed. For the model with $m_{\rm
  \nu}=0.5$ keV, the free-streaming mass is about $10^{9}M_{\odot}$.
Actually also the formation of halo more massive than this cut-off can
also be affected up to a mass of the order of $2000\times M_{\rm fs}$
(e.g. Schneider et al. 2012).

In Fig. \ref{fig:HMFs} we show the dark matter halo
mass functions for our different models.  Note that this mass function
is slightly different from the usual mass function that normally only
includes Friends Of Friends (FOF) haloes (i.e. central haloes). In our
mass function we also include satellite halos: their mass is defined
as the mass at the accretion time $M_{\rm acc}$, in other words the
mass the satellite had before becoming a satellite of a more massive
halo.  It has been shown that this accretion mass is more closely related
to the final stellar mass of the satellite (e.g., Vale \& Ostriker
2004, Moster \etal 2010).  For central halos, the mass is simply
defined as the virial mass at the current time.  Such a mass function
which also includes satellite galaxies can be easily linked to the
predicted galaxy mass functions, which is one of the observations we
want to study in Warm Dark Matter.

For our more extreme WDM model with $m_{\rm \nu}=0.5$ keV, the halo
number density is lower by roughly 60\% at $M_{\rm DM}=10^{11.5}
M_{\odot}$.  This confirms that the free-streaming in WDM has
non-negligible effects on the formation of haloes with mass much
larger than $M_{\rm fs}$.  On the other hand the model with $m_{\rm
  \nu}=2$ keV, has a cutoff mass of $M_{\rm fs} \approx
10^{6}M_{\odot}$ which is well below the mass resolution of our
simulations, as a consequence we do no see any differences with
respect to the CDM model on the whole mass range probed by our runs.
The warmest model also presents an upturn in the halo mass function
around $M_{\rm DM} < 10^{11.5}M_{\odot}$.  This upturn is due to the
artificial clumping effect in WDM simulations from the limited
resolution (Wang \& White 2007; Schneider \etal 2012), which indicates
that low mass haloes are possibly over-predicted in our
simulations. We will later briefly discuss the implication of this
effect on the prediction of galaxy properties.  Finally all model
agree reasonably well for masses higher than $\sim 10^{12.5}
\Msun$, so we do not expect large differences for the properties of
galaxies living in massive haloes.

By applying the Semi Analytical Model to the N-body simulations it is
possible to compute properties of the galaxies harbored in the dark
matter haloes in the various WDM and CDM models.
Fig.~\ref{fig:SMFs} shows the stellar
mass functions from $z=0$ to $z=3.5$. In this first
attempt we apply the same SAM (with the same parameters) to all the
models.

At $z=0$ the WDM model with lower $m_{\rm \nu}$ produces too few
galaxies around $M_{\rm star}= 10^{10}M_{\odot}$ with respect to
current observational data; this is a direct consequence of the lower
normalization to the halos mass function below $M_{\rm DM}=
10^{11.5}M_{\odot}$ in WDM with respect to CDM.  The model for 0.75
keV gives a stellar mass function between the previous model and the
CDM prediction, while the model with $m_{\nu}=2$ keV produces almost
identical results to the CDM model (as expected from the results on
the halo mass function) and we will not show its results in the rest
of the paper.
The fact that WDM models seem to produce too few galaxies at low
stellar masses with respect to observations does not come as a
surprise.  The parameters in the SAM we use have been {\it tuned} to
reproduce the stellar mass function at $z=0$ in the (standard) CDM
scenario, so any models that predict less low mass dark matter haloes
will, by construction, underpredict the number of galaxies at low
stellar mass.

Results are somewhat inverted at higher redshift. The CDM model overpredicts the
number of galaxies with a stellar mass below $10^{10.5} \Msun$. This
is a well know problem of the current modeling of galaxy formation
(e.g., Fontanot \etal 2010, see Weinmann \etal 2012 for a thorough
description of this issue).  In this case having a warm dark matter
component naturally helps in reducing the number of low mass haloes,
and consequently the number of low (stellar) mass galaxies, bringing
SAM results in better agreement with observations.  But of course this
``success'' comes at the sacrifice of local ($z=0$) agreement.

Recently Menci \etal (2012) used this suppression of stellar mass
function at high redshift in warm dark matter model as a hint that WDM
may help in reducing the tension between data and theoretical
prediction at high redshift.  In their case the adopted SAM was also
overpredicting the luminosity function at $z=0$ for the CDM model,
where on the contrary our SAM+CDM model provides a good fit to the
local data. So having a WDM component in the case of Menci \etal
(2012) was helping on both high and low redshift.

What remains to be clarified is whether the success of WDM claimed by
these authors is truly due to the WDM component or simply reflects a
non-optimal parameterization of the physics of galaxy formation
implemented in their SAM (that could  partially be the
explanation of the failure of the CDM scenario in their model 
at both high and low luminosities, at $z=0$ and $z=1.5$).

\begin{figure*}
 \centerline{\psfig{figure=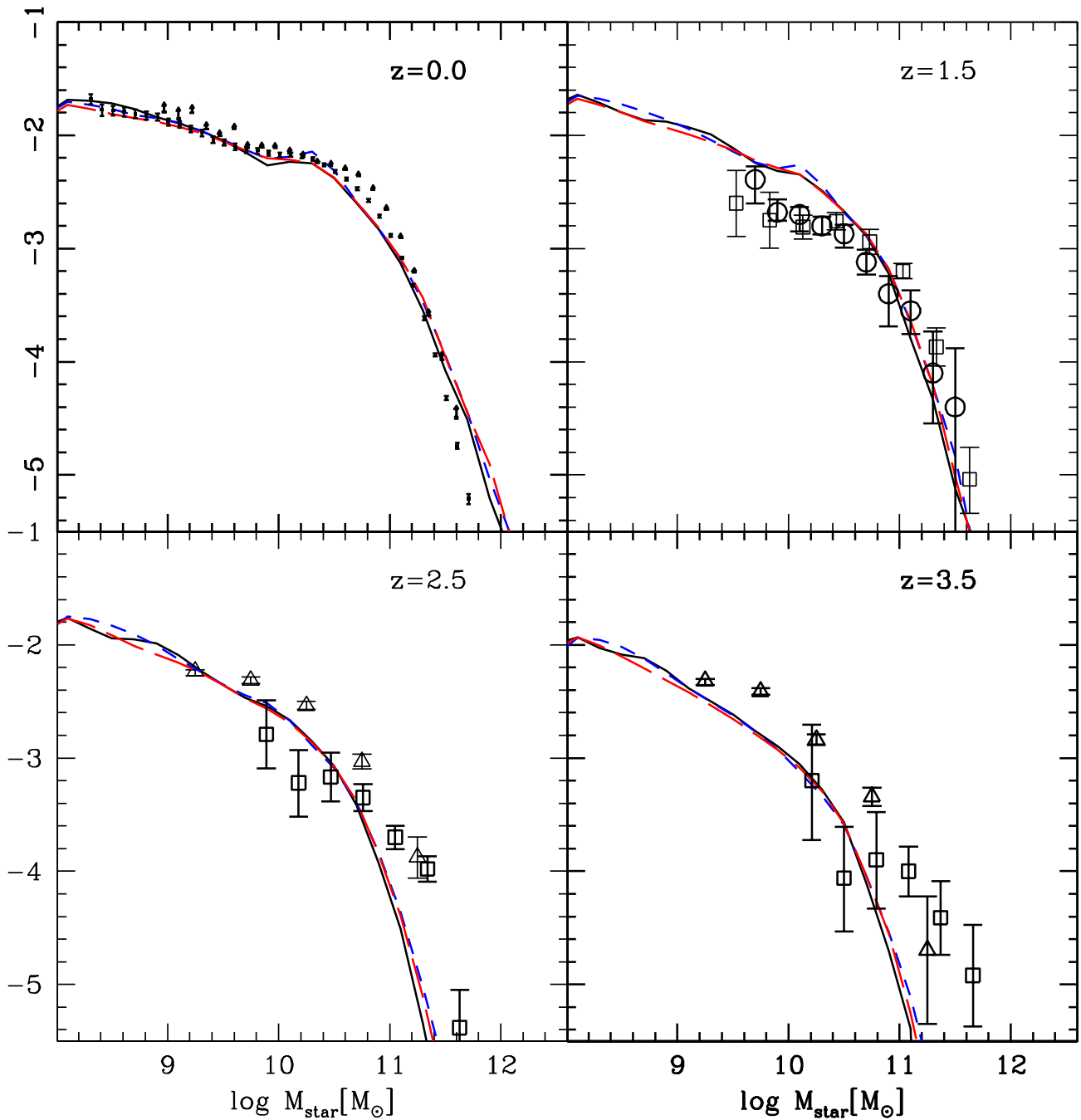,width=0.95\textwidth}}
 \caption{The galaxy stellar mass functions. As in Fig.~\ref{fig:SMFs}, the points show the
   observations and the lines show the models.  For the WDM models the
   cooling factor is treated as a free parameter.  WDM models with
   $m_{\rm \nu}=0.75, 0.5$ keV, have $f_{\rm c} =0.35, 1.0$
   respectively.}
\label{fig:MFnew}
\end{figure*}

In the same paper Menci and collaborators also found an effect of WDM
on the high luminosity tail of the luminosity function. Our results do
not show such an effect -- all our models give the exact same
prediction at high stellar masses. This result is somehow expected
given the similarity of the halo mass function at high masses (see  Fig.  ~\ref{fig:HMFs}). Menci \etal claimed that this
difference was due to the different accretion rates of satellites in
massive haloes.  SAMs have been show to have problems in modelling the
properties of satellites in high mass haloes (e.g., Weinmann \etal
2010, Liu \etal 2010 and references therein).  This again raises the
question whether the effect seen by Menci and collaborators (but not
in this work) is a true effect due to warm dark matter, or simply
reflects the specific way in which baryonic physics is implemented in
their semi analytical model.  We will address this issue of degeneracy
between SAM parameters and DM properties in the next section.

\subsection{Degeneracy between SAM parameters and WDM}

The failure of WDM models in matching the redshift zero stellar mass
function does not imply that WDM models are somehow ``wrong''; the
reason for this resides in the large uncertainty of baryonic physics
parameterization in the SAMs.  Obviously the same argument applies to
the ``success'' of WDM models in reproducing the observational data at low-mass end at high redshift.

Fig.~\ref{fig:SMFs} suggests that the stellar mass
in low-mass halo is too low in WDM models.  One possible solution
would be to increase the star formation efficiency or decrease the
effect the supernova feedback in the model, but this affect the high
mass tail of the stellar mass function, where all models perform
equally well.

Instead we decided to turn to the effective cooling factor given in
eq.~\ref{eq:cooling}, since this factor is introduced only for
low-mass haloes.  To get a better match to the local stellar mass
function, we set $f_{\rm c}=1.0$ and $0.3$ in the WDM models with
$m_{\nu}=0.5$ and $0.75$ keV, respectively.

Fig.~\ref{fig:MFnew} shows the model predictions for these new values
of $f_{\rm c}$. With this new effective cooling factor both WDM
models (0.5 and 0.75 keV) provide equally good fit to the data at
$z=0$ as the CDM model.  On the other hand they now show the same
over-prediction of galaxies at $z>0$ as CDM. This is due to the very
similar evolution of WDM and CDM models (see Fig.~1 in Weinmann \etal 2012).

Overall we find that by simply tweaking one single parameter we have
been able to get the exact same predictions in WDM and CDM
models. This shows the danger in using a single observable (e.g. the
stellar mass function) to claim the success of one dark matter model
over another.
In the next section we will try to combine different observables to
break the degeneracy between the DM properties and the parameterization
of baryonic effects in SAMs.

\subsection{Constraining WDM with the stellar-to-halo mass ratio}

In Fig.~\ref{fig:MFnew} we have shown that better match to the local
stellar mass function can be achieved in the WDM models by increasing
the $f_{\rm c}$ parameter in low-mass haloes.  Of course this has the
consequence of increasing the stellar mass content in low mass haloes.

This increased ratio between stellar and halo mass can be constrained
using several independent observations. The halo mass can be directly
measured for stacks of galaxies using galaxy-galaxy weak lensing
(e.g., Mandelbaum \etal 2006; Leauthaud \etal 2012a,b) or satellite
kinematics (e.g., Conroy \etal 2007; More \etal 2011), and inferred
assuming halo abundance matching (Moster et al. 2010).
A closely related quantity to the halo mass is the maximum circular
velocity of the dark matter halo. The best observational probe of this
is the rotation velocity in the ``flat'' part of the rotation curve,
$V_{\rm flat}$, as traced by 21cm observations of neutral hydrogen. We
refer to the relation between stellar mass and outer rotation velocity
as the Tully-Fisher relation, even though the original relation from
Tully \& Fisher (1977) was between B-band luminosity and 21cm
linewidth.

\begin{figure}
 \centerline{\psfig{figure=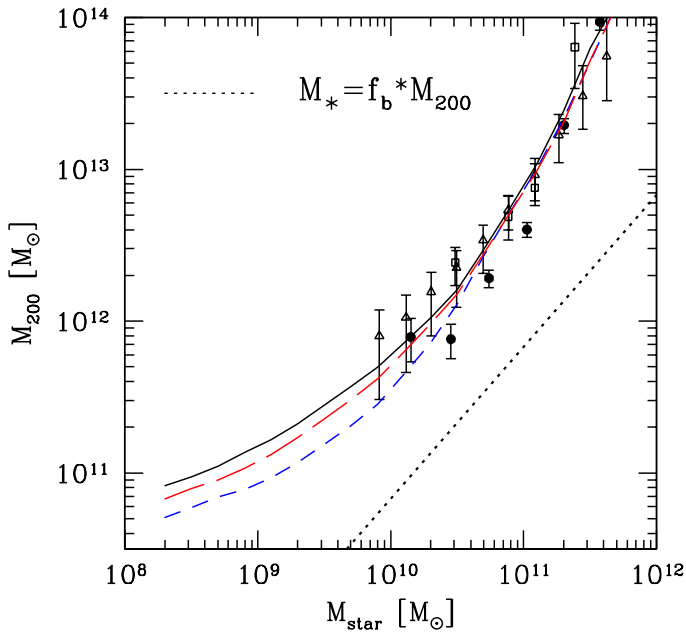,width=0.5\textwidth}}
 \caption{The stellar mass to halo mass relation. The data points are
   observational results using different approaches to determine the
   halo mass: Galaxy-galaxy weak lensing from Mandelbaum \etal 2006
   (filled circles); Satellite kinematics from Conroy \etal 2007 (open
   squares) and More \etal 2009 (open triangles).  The lines are the
   predictions from the SAMs (CDM -- solid black; WDM 0.75 KeV -- red
   long-dashed; WDM 0.5 KeV -- blue short-dashed).  The dotted line
   shows the maximum stellar mass in given halo mass assuming an
   universal baryonic fraction.}
\label{fig:Ms-Mh}
\end{figure}

In Fig.~\ref{fig:Ms-Mh} we show the halo mass to stellar mass relation
at $z=0$ for our WDM models (with different $f_{\rm c}$) and the CDM
model.  Data points with error-bars are measured stellar to halo mass
ratio from weak lensing and satellite kinematics (see Leauthaud \etal
2012a for a discussion of the different data sets.).  The dotted line
shows the maximum stellar mass in given halo mass assuming the
universal baryon fraction, $f_{\rm b}=\Omega_{\rm b}/\Omega_{\rm m}$
from the WMAP7 cosmology.

While the three models give the same stellar mass function (as shown
in Fig. \ref{fig:MFnew}), they do predict different halo-to-stellar
mass ratios especially for halo masses below $10^{12}M_{\odot}$.  For
a give halo mass the WDM models predict a large stellar mass with this
difference increasing for decreasing $m_{\nu}$ values. Unfortunately,
the current weak lensing and satellite kinematics observations do not
probe halo masses below $10^{12}M_{\odot}$.

\subsection{Constraining WDM with the Tully-Fisher relation}
\begin{figure}
 \centerline{\psfig{figure=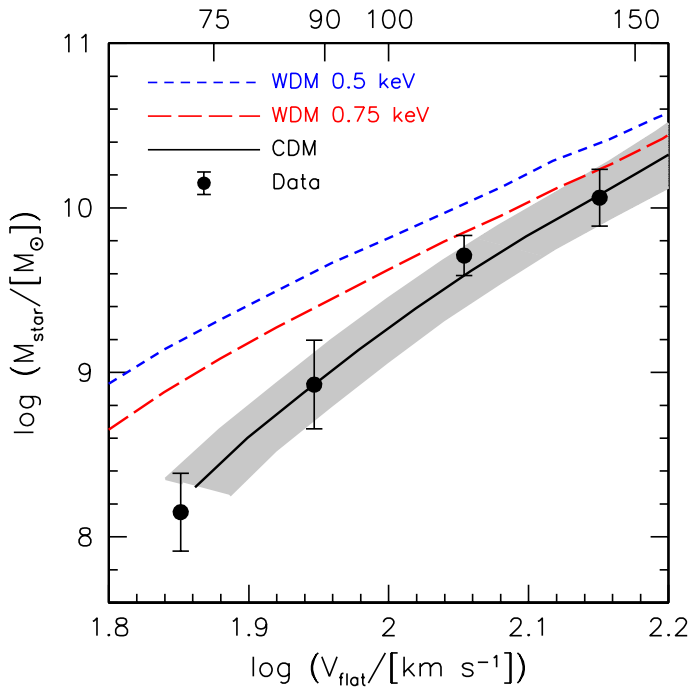,width=0.5\textwidth}}
 \caption{The Tully-Fisher relation. The lines show the same models as
   in Figs.~\ref{fig:MFnew}\&\ref{fig:Ms-Mh}. For the models we
   calculate $V_{\rm flat}$ at a radius enclosing 80\% of the cold
   gas.  The filled circles with error bars show the mean (and error
   on the mean) of $\log M_{\rm star}$ in bins of $V_{\rm flat}$ using
   observations from Stark \etal (2009) and McGaugh (2012). The shaded region shows the effect of halo contraction and expansion. The CDM
   model is clearly favored over both WDM models.}
\label{fig:TF}
\end{figure}

For haloes with masses in the range $10^{11}< M_{200}/M_{\odot} <
10^{12}$, the best current probe of the halo masses comes from the
Tully-Fisher relation. This constraint is shown in Fig.~\ref{fig:TF}.
As before the observations are shown with points and error bars, while
the models are shown with lines. For the observations the points show
mean of $\log M_{\rm star}$ in bins of outer rotation velocity, with
the error bar corresponding to the error on the mean. There are
typically $\sim 10$ galaxies per velocity bin. The observations are
from the compilations of Stark \etal (2009) and McGaugh (2012). We
calculate stellar masses using relations from Bell \etal (2003) (with
a $-0.1$ dex correction to convert to a Chabrier 2003 IMF). We also
convert the Hubble parameter from 75 to 70.

For the models we compute the circular velocity at a radius enclosing
80\% of the cold gas, $R_{80}$. At this radius model rotation curves
tend to be flat (Dutton 2009).  The SAM does not provide radial
information for the stars, gas, or dark matter, so we use empirical
constraints (from Dutton \etal 2011) for the stars and theoretical
constraints (Maccio \etal 2008; Schneider \etal 2012) for the dark
matter.
We assume the stars and cold gas are in exponential disks, with the
scale length of the cold gas being 1.55 times larger than that of the
stellar disk (Dutton \etal 2011). We determine the average disk size
using the size vs stellar mass relation for late-type galaxies from
Dutton \etal (2011).  Thus for a given stellar mass in the model, we
know the average radius we should be measuring the rotation velocity,
and the contribution of the stars and cold gas to this velocity.
For the dark matter and hot gas we assume the profiles are NFW. For
the CDM case we adopt the concentration mass relation from Macci\`o
\etal (2008) for a WMAP5 cosmology (which is very similar to that of
the cosmology adopted here).
For the WDM models we scale the concentration parameters according to
the fitting formula of Schneider \etal (2012). We also consider two 
possiblities for the halo response to the baryonic mass, either by contraction (Gnedin et al. 2004) or expansion (Dutton et al. 2011). Their effects are shown as the shaded region in Fig.~\ref{fig:TF}.

Fig.~\ref{fig:TF} shows that the CDM model reproduces the data well,
whereas at a fixed velocity the WDM models have higher stellar masses,
especially at low velocities. At a velocity of $V_{\rm flat}\sim 80
{\rm km s^{-1}}$ the differences are substantial: a factor of $\approx
2$ between CDM and 0.75 keV WDM, and a factor of $\approx 1.6$ between
0.5 and 0.75 keV WDM. These large differences are due to two
effects. Firstly, as shown in Fig.~\ref{fig:Ms-Mh} at fixed halo mass
WDM models have higher stellar masses. Secondly, WDM models have lower
concentration haloes, which results in lower rotation velocities, and
hence higher stellar mass at fixed velocity. We found that 
the second effect is the dominant one, as in the first case  simply increasing the stellar will also increase the circular velocity.

In summary the CDM model provides a good match to the data (note that
the model was {\it not} tuned to match the TF relation), whereas both
both WDM models overpredict the stellar masses at fixed velocity.  The
limit of $m_{\nu} > 0.75$ keV that we find here is consistent with
current limits from large scale structure (e.g. Viel \etal 2008).

\section{Conclusions}
\label{sec:cons}

Recent observational results have challenged the otherwise successful
Cold Dark Matter model on small scales. For example the inner density
profile of dark matter halo is too concentrated to match the
kinematics of satellites around the Milky Way (e.g. Boylan-Kolchin
\etal 2011).
Warm Dark Matter models have been suggested as a possible solution
since the suppression of power on small scales can in principle
alleviate if not solve these issues (e.g. Lovell \etal 2012).

However, Most of the CDM predictions are based on gravitational only
(N-body) simulations, which by construction neglect the effects of
baryons and their complicated network and interactions through gas
cooling, star formation and feedback that could possibly alter the
results of pure DM simulations (e.g. Governato \etal 2012, Brooks \etal 2012)
It is then important to investigate the prediction of CDM and WDM on
the statistical properties of galaxy population, moving beyond a
simple DM simulation.

In this paper, we performed high-resolution N-body
cosmological simulations for three WDM models with $m_{\rm \nu}=0.5$, 0.7
and 2.0 keV respectively, plus a {\it controlled} CDM model. We couple
these simulations with a Semi Analytical Model of galaxy formation to
study the impact of a cut-off in the power spectrum on observable
quantities such as the stellar mass function.

For a fixed set of parameters describing the baryonic physics, models
with low masses for the warm particle ($m_{\nu}=0.5$ and 0.75 keV),
predict less galaxies with stellar mass $M_{\rm star}
<10^{10}M_{\odot}$ than the current data at $z=0$.  We find that the
WDM model with $m_{\rm \nu}=2$ keV provides almost identical results
as CDM and it is able to successfully reproduce the data. The
situation is reversed at higher redshift $z=1.5$, with WDM models in
better agreement with the observed stellar mass function than the CDM
model (or the WDM model with $m_{\rm \nu}= 2$ keV).

However we show these differences in the stellar mass function, are
strongly degenerate with the set of parameters used in the Semi
Analytical Models to describe the (largely unknown) galaxy formation
processes.  By adjusting a single parameter (the cooling rate of low
mass haloes) we have been able to get indistinguishable results from
all the DM models, both at high and low redshift.  This shows that a
single observable (e.g. the stellar mass function) can not constrain
the effects of warm component on the galaxy formation process.

In order to break this degeneracy we use independent constraints on
the (integrated) star formation efficiency at low masses.  We found
that if the WDM models are tuned to  reproduce the present
($z=0$) stellar mass function, the stellar mass for a given halo mass
is systematically larger in haloes of masses below ($\sim
10^{12}M_{\odot}$). Such scales are not reliably probed by direct
methods such as satellite kinematics or weak galaxy-galaxy lensing.
In order to probe such low halo masses we use the relation between
stellar mass and rotation velocity at large galactic radii, more
commonly known as the Tully-Fisher relation. Current data already rule
out models with $m_{\rm \nu}<0.75$ keV which is in agreement with
other limits from large scale structure (e.g. Viel \etal 2008).

Finally our study shows that by combining measurements of galaxy
stellar mass function and stellar mass - halo mass relation down to
low galaxy masses ($\approx 70$km/s) it is possible to obtain very
tight constraints on the mass of a possible warm component.  This opens a
new window in the search for the nature of the elusive dark matter
component of our Universe.

\acknowledgments

We thank Ming  Li for helping us running the WDM simulations and Alexie Leauthaud
for  kindly  providing the  compiled  data  points  in Fig.3.   XK  is
supported by  the Bairen program  of the Chinese Academy  of Sciences,
the foundation for the  author of CAS excellent doctoral dissertation,
and  NSFC (No.  11073055). The  simulation runs  are supported  by the
Supercomputing center of CAS.
The authors acknowledge support from the MPG-CAS through the partnership program
between the MPIA group lead by AVM and the PMO group lead by XK.
AVM and AD AVM acknowledge 
support by Sonderforschungsbereich SFB 881 ‘The Milky Way System’ (subproject A1) 
of the German Research Foundation (DFG).


\end{document}